# Grading Practices and Considerations of Graduate Students at the Beginning of their Teaching Assignment

Edit Yerushalmi[1], Emily Marshman[2], Alexandru Maries[2], Charles Henderson[3] and Chandralekha Singh[2]

[1] Department of Science Teaching, Weizmann Institute of Science, 234 Herzl St., Rehovot, Israel 7610001
[2] Department of Physics and Astronomy, University of Pittsburgh, 3941 O'Hara St., Pittsburgh, PA 15260
[3] Department of Physics, Western Michigan University, 1903 W. Michigan Ave., Kalamazoo, MI, 49008

**Abstract:** Research shows that expert-like approaches to problem-solving can be promoted by encouraging students to explicate their thought processes and follow a prescribed problem-solving strategy. Since grading communicates instructors' expectations, teaching assistants' grading decisions play a crucial role in forming students' approaches to problem-solving in physics. We investigated the grading practices and considerations of 43 graduate teaching assistants (TAs). The TAs were asked to grade a set of specially designed student solutions and explain their grading decisions. We found that in a quiz context, a majority of TAs noticed but did not grade on solution features which promote expert-like approaches to problem-solving. In addition, TAs graded differently in quiz and homework contexts, partly because of how they considered time limitations in a quiz. Our findings can inform professional development programs for TAs.

## INTRODUCTION

Problem-solving (PS) plays a central role in physics teaching. Research has shown that instruction can promote expert-like approaches to PS by encouraging students to follow a prescribed PS strategy that explicates the tacit PS processes of an expert, [1] including: 1) describing the problem situation in physics terms; 2) planning the construction of a solution; and 3) evaluation. Research has also shown that instruction can foster learning domain knowledge through PS by encouraging students to articulate their reasoning, reflect and self-explain how domain concepts and principles were applied to solve a problem, acknowledge differences between their own and others' approaches to a problem, and attempt to resolve arising conflicts [2]. Thus, within an instructional approach based on formative assessment [3], grading should reward explication of reasoning and the use of a prescribed PS strategy.

A central way to influence grading practices in a physics classroom is through graduate TAs, both because TAs are often responsible for grading students' work and because TAs are often required to participate in professional development (PD) programs. PD should be based on research about the beliefs and practices of TAs. As one piece of this research, we studied 43 graduate TAs enrolled in a PD program at the University of Pittsburgh. In this context, we investigated: What are TAs' grading practices? Which features do they consider when grading? What are their reasons for weighing solution features?

## METHODOLOGY

Data collection took place at the beginning of the TAs' teaching career, within the first month of a PD program conducted by a PER faculty member during the fall semester. TAs filled out a worksheet designed to encourage introspection regarding instructional choices related to grading [4, 5]. The worksheet asked TAs to make judgments about a set of solutions designed to reflect both common student responses to a context-rich physics problem (see Fig. 1) as well as expert-like and novice approaches. Here we focus on two of the five solutions (see Fig. 2). Clearly incorrect aspects of the solutions are indicated by boxed notes. The TAs graded the student solutions for both homework and quiz contexts. For each solution, they were asked to list characteristic features and explain how and why they weighed those features to obtain a score (see Fig. 3).

> You are whirling a stone tied to the end of a string around in a vertical circle having a radius of 65 cm. You wish to whirl the stone fast enough so that when it is released at the point where the stone is moving directly upward it will rise to a maximum height of 23 m above the lowest point in the circle. In order to do this, what force will you have to exert on the string when the stone passes through its lowest point one-quarter turn before release. Assume that by the time that you have gotten the stone going and it makes its final turn around the circle, you are holding the end of the string at a fixed position. Assume also that air resistance can be neglected. The stone weighs 18 N.

**FIGURE 1.** Problem Statement

We suggest that the reader examine the student solutions and think about how to grade them.

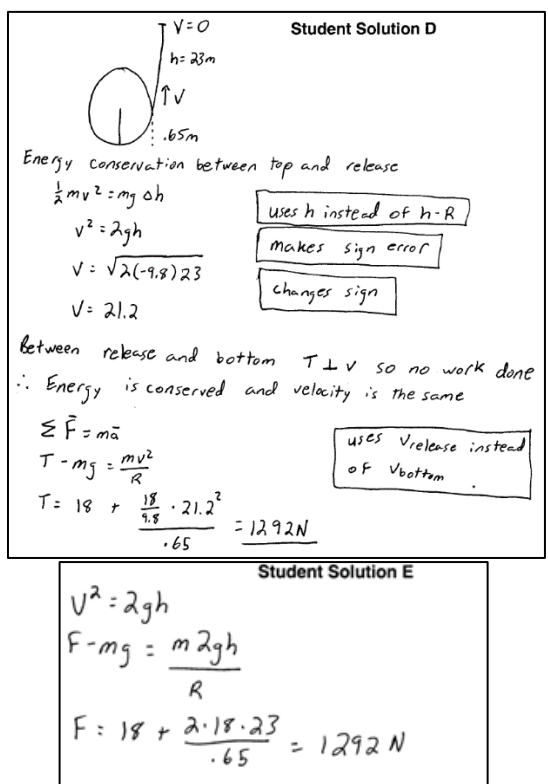

**FIGURE 2.** Student Solution D (SSD) and E (SSE)

| Features: Solution E | Score | | Reasons: Explain your weighing of the different features to obtain a score |
|---|---|---|---|
| | Q | HW | |
| No word explanation No figure No error Precise and concise | 10 | 9 | *There are no explanations in this solution, which means I could not know whether the student really knows the process or he/she just misdid like solution D. This is why I put 1 point off from this solution if this was HW. However, in the quiz time is limited, I will give a full grade to this solution* |

**FIGURE 3.** Sample TA worksheet for SSE.

The student solutions were designed to reflect expert and novice approaches to PS and to trigger conflicting instructional considerations in assigning a grade. For example, in comparing solution SSD to SSE, note that both include the feature of a correct answer. However, only SSD includes a diagram, articulates the principles used to find intermediate variables, and provides clear justification for the final result. In contrast, SSE is brief with no explication of reasoning. However, the elaborated reasoning in SSD reveals two canceling incorrect calculations, involving misreading of the problem situation as well as misuse of energy conservation to imply circular motion with constant speed. In contrast, SSE, being very brief, does not give away any evidence for mistaken ideas, even though the student might be guided by a similar thought process as Student D. Thus, TAs' grading of SSE and SSD could reveal to what extent they encourage the use of a prescribed PS strategy and showing reasoning explicitly.

Data analysis involved coding the solution features listed by TAs in the worksheets (see Fig. 3) into a combination of theory-driven and emergent categories. The features were also coded for whether they were merely mentioned or weighed in grading. For example, the sample TA listed "no figure" as a feature in SSE, but when assigning a grade, did not refer to this feature when explaining how s/he obtained a score. We identified 21 features that were grouped into 5 clusters.

As shown in Table 1, cluster 1 (C1) included both features related to initial problem analysis as well as evaluation of the final result. C2 involves features related to explication of reasoning (i.e., articulation and justification of principles). We consider that TAs who grade on C1 and C2 are encouraging students to follow a prescribed PS strategy. C3 includes domain-specific features, such as invoking relevant physics concepts and principles and applying them properly. C4 includes features related to elaboration which emerged during the coding process. These features were not assigned to the "explication" category as they were imprecise (e.g., "written statements" could be interpreted to mean articulation of principles or simply a written explanation of the physical setup). Features in C4 could be productive, counterproductive, or neutral in encouraging expert-like PS approaches (assigned +, -, 0 respectively). For example, grading for conciseness could transmit a message to the students that physics problems should be solved with little detail (assigned as (-) for being counterproductive), while grading for written statements could transmit a message that explication of the thought process is important for learning from PS (assigned (+) for being productive). Finally, C5 focuses on correctness of algebra and final answer. TAs who give a large weight to these features may transmit a message to the student that the final result is acceptable without justification.

**TABLE 1.** Sample features sorted into clusters

| C1 Problem description & evaluation | | Visual representation; articulating the target variables and known quantities (e.g, "knowns/ unknowns"); evaluation of the reasonability of the final answer (e.g., "check") |
|---|---|---|
| C2 Explication of PS approach | | Explicit sub-problems (e.g., "solution in steps"); articulation of principles (e.g., "labels energy conservation use"; justifying principles (e.g., "explained the reason he used the formulas") |
| C3 Domain knowledge | | Essential principle invoked (e.g., "sums forces, energy conservation") ; essential principle is applied adequately |
| C4 Elaboration | + | Explanation; written statements |
| | 0 | Organization; showing algebraic steps |
| | - | Conciseness |
| C5 correctness | | Algebraic errors; correct final answer |

# RESULTS

## Grading Practice

We found that in a quiz context, TAs graded a solution which provides minimal reasoning while possibly obscuring physics mistakes (SSE) higher than a solution which shows detailed reasoning and includes canceling physics mistakes (SSD). In the quiz context, many more TAs graded SSE>SSD (N=28, 65%) compared to SSD>SSE (N=10, 23%), transmitting a message that is counterproductive to promoting the use of prescribed PS strategies and providing explication of reasoning. We found a similar gap in the HW context, although the gap is somewhat softened: 58% of TAs (N=25) graded SSE>SSD while 35% (N=15) graded SSD>SSE. In a quiz context, TAs graded SSE significantly higher than SSD (<SSE>=8.3 compared to <SSD>=7.1, p-value calculated by a t-test: 0.010) while in a homework, the averages are comparable (<SSE>=7.1 and <SSD>=6.7).

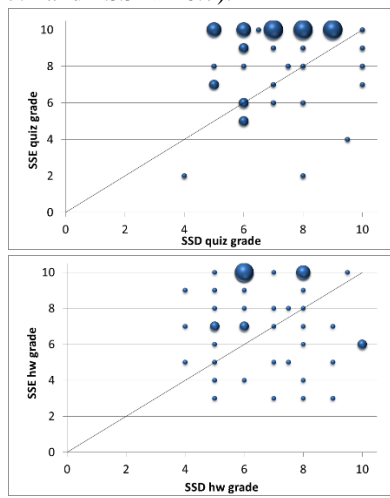

**FIGURE 4.** Distribution of TA grades, quiz and HW.

## Features considered in grading

In order to quantitatively represent the features weighed by groups of TAs who are likely to have differing considerations in grading, we display the distribution of features mentioned and graded on by TAs who graded SSE>SSD and TAs who graded SSD>SSE, overlooking SSE=SSD. These distributions for the quiz (Q) and homework (HW) contexts vary depending on the solution as shown in Table 2.

We found a significant gap between the percentage of TAs who mentioned features from clusters which promote prescribed PS strategies and the percentage of TAs who graded on these features. This gap is more evident in the SSE>SSD group, in the quiz as well as in the HW context.

**TABLE 2.** Feature distribution for quiz and homework. Bold italics indicate ~50% or more TAs grade on the cluster.

| Cluster | | SSE>SSD group (Q: N=28, HW: N=25) | | | | SSD>SSE group (Q: N=10, HW: N=15) | | | |
|---|---|---|---|---|---|---|---|---|---|
| | | Mention % | | Grade % | | Mention % | | Grade % | |
| | | Q | HW | Q | HW | Q | HW | Q | HW |
| C1 | SSE | 46 | 36 | 7 | 4 | 60 | 73 | 20 | 13 |
| | SSD | 43 | 40 | 4 | 8 | 80 | 73 | 20 | 20 |
| C2 | SSE | 32 | 32 | 11 | 24 | 10 | 27 | 10 | 27 |
| | SSD | 50 | 48 | 25 | 24 | 40 | 53 | 30 | 27 |
| C3 | SSE | 14 | 12 | 11 | 8 | 40 | 20 | 40 | 13 |
| | SSD | 89 | 92 | ***79*** | ***80*** | 70 | 73 | ***70*** | ***73*** |
| C4 (+) | SSE | 46 | 48 | 18 | 28 | 80 | 60 | ***60*** | ***53*** |
| | SSD | 18 | 20 | 7 | 8 | 40 | 27 | 20 | 13 |
| C4 (0) | SSE | 39 | 32 | 7 | 20 | 50 | 60 | 30 | 33 |
| | SSD | 18 | 12 | 11 | 8 | 20 | 33 | 10 | 20 |
| C4 (-) | SSE | 32 | 32 | 14 | 8 | 0 | 13 | 0 | 0 |
| | SSD | 4 | 0 | 4 | 0 | 0 | 7 | 0 | 7 |
| C5 | SSE | 43 | 40 | 14 | 8 | 50 | 40 | 20 | 13 |
| | SSD | 75 | 76 | ***43*** | ***52*** | 80 | 67 | ***70*** | ***53*** |

Regarding cluster C1, (problem description and evaluation), 20% or less of the TAs stated that they grade on these features in both SSE and SSD. Also, slightly more TAs who graded SSD>SSE than TAs who graded SSE>SSD considered C1 when grading (13%-20% compared to 4%-8%). Many more TAs mentioned this cluster even though they did not consider it in their grading (46% in SSE>SSD group, 80% in SSD>SSE group). We conclude that even though TAs mentioned the cluster of problem description and evaluation, they refrained from grading on it regardless of whether it is missing (as in SSE) or present (as in SSD).

Regarding cluster C2, which involves explication, there is a lot of similarity between the SSD>SSE and the SSE>SSD groups: both refrained from grading on this cluster in the quiz context (~10%) or in the HW context (~25%) for SSE. A larger portion of TAs stated that they grade on this cluster in SSD (25%-30%) than in SSE (10%-11%) on a quiz. Similar to C1, many more TAs mentioned this cluster even though they did not consider it in their grading.

Cluster C4+ relates also to explication, however, in an ill-defined manner (see Table 1). Similar to C1 and C2, many more TAs noticed features from C4+ than graded on these features. However, the difference between the two groups becomes more prominent. In the SSD>SSE group in the quiz (60%) as well as HW (53%) context more than half of the TAs graded on this cluster in SSE, while much fewer graded on it in the SSE>SSD group (18%-28%). When grading SSD fewer TAs graded on this cluster.

This last result can be interpreted to indicate that TAs may use a subtractive grading scheme, taking points from SSE for missing explanations (C4+), but not weighing this feature in grading SSD, where it is represented. Using a subtractive grading scheme is evident also from analyzing other clusters that are most prominent in TAs' grading: domain knowledge (C3) and correctness (C5) (see italicized percentages in Table 2). Over 70% of all TAs graded on physics knowledge in SSD, where physics concepts and principles are inadequately applied. However, fewer TAs said that they grade on domain knowledge in SSE. Additionally, ~50% of all TAs graded on correctness (errors) in SSD and few (less than 20%) on SSE.

## Reasons for grading

We noted previously that we found a difference in TAs' grading practices in the HW and quiz contexts, where TAs are more inclined to insist on explication of reasoning in the HW as compared to the quiz context. However, we did not find significant differences in TAs' grading of clusters in these two contexts. To understand this discrepancy, we examined TAs' reasons for weighing different solution features (listed in the right hand column in the worksheet they completed, see Fig. 3). We focus here on TAs' grading of SSE in a quiz context. The reasons were coded in a bottom-up manner, resulting in the four categories shown in Table 3. Table 3 shows that the difference in grading may stem from TAs' consideration of evidence of students' thought processes, consideration of time limitations in a quiz, or their preference for aesthetics (physics problems should be solved in a brief, condensed way).

**TABLE 3.** Reasons for SSE grade in quiz. $n$ - number of TAs that provided reasons out of N (number of TAs in each group). Each TA could provide more than one reason.

| Reasons | SSE>SSD (N=28 total, $n$=16) | SSD>SSE (N=10 total, $n$=6) |
|---|---|---|
| Adequate evidence | 7 | 0 |
| Inadequate evidence | 3 | 6 |
| Time/stress | 5 | 0 |
| Aesthetics | 5 | 0 |

## DISCUSSION AND SUMMARY

An analysis of TAs' grading practices and considerations at the beginning of their teaching assignment reveals the following:

- In a quiz context, a majority of TAs gave a higher grade to a solution that provides minimal reasoning while possibly obscuring physics mistakes as compared to a solution providing reasoning that reveals canceling mistakes. Their grading did not encourage students to use prescribed PS strategies nor to explicate their reasoning process. This tendency softened somewhat in a homework context.
- While TAs' grading differs in quiz and HW context, there is little difference in the solution clusters they considered in both contexts. Many TAs were aware of features related to explication and prescribed PS strategies, but few graded on these same features.
- TAs' grading approaches indicate that they used a subtractive scheme. Most TAs considered domain knowledge (C3) and correctness (C5) to a larger extent than other clusters and when using a subtractive scheme, they often only recognized errors rather than missing justifications. In turn, their grading may transmit a message that explication of problem description, planning of the solution, and evaluation are not required in students' solutions.
- The difference between HW and quiz grading may stem from how TAs considered time limitations in a quiz as a reason for accepting brief answers as adequate evidence of students' thought processes.

The results of this study concerning TAs' grading practices and underlying considerations are consistent with prior work on TAs' practices and considerations when designing example solutions for students [5]. In Ref. [5], the majority of TAs' own solutions included neither explication of reasoning nor a reasonability check of the final answer. Similarly, in the grading study, very few TAs grade on articulation and justification of principles (C2) and checking of the final answer (C1). This suggests that TAs neither design example solutions nor grade student solutions in a manner which promotes the use of expert-like PS strategies to help students learn from PS.

Since this investigation took place at the beginning of the TAs' teaching career, the results can serve to inform PD activities to prepare TAs for their grading responsibilities. As in other learning environments, PD should also elicit TAs' ideas and allow them to reflect and try to resolve conflicting ideas and approaches to physics instruction. The conflicts between the features that TAs are aware of, the features that they grade on, and the differences in how TAs consider adequate evidence of students' thought processes in different settings could serve as fruitful starting points for such discussions. In this way, TAs can be guided to implement grading practices that promote the development of expert-like approaches to PS.